\documentclass[12pt]{iopart}
\usepackage{graphicx}
\usepackage{dcolumn}
\usepackage{bm}

\begin{document}

\title{Logical operations with Localized Structures}

\author{Adrian Jacobo \footnote{Present address: Max Planck Institute for the Physics of Complex Systems, N\"{o}thnitzer Stra{\ss}e 38, 01187 Dresden, Germany}}
\author{Dami\`a Gomila}
\ead{damia@ifisc.uib-csic.es}
\author{Manuel A. Mat\'ias}
\author{Pere Colet}
\address{IFISC, Instituto de F\'{\i}sica Interdisciplinar y Sistemas Complejos (CSIC-UIB),E-07122 Palma de Mallorca, Spain}
\date{\today}
\submitto{\NJP}

\begin{abstract}
We show how to exploit excitable regimes mediated by localized structures (LS) to
perform AND, OR, and NOT logical operations providing full logical functionality. Our scheme is general and can be implemented in any physical system displaying LS. In particular, LS in nonlinear photonic devices can be used for all-optical computing applications where several reconfigurable logic gates can be implemented in the transverse plane of a single device, allowing for parallel computing.
\end{abstract}


\maketitle
\section{Introduction}
Electronic computers are prevalent around us, and have been immensely successful in shaping our technological society. Most practical computers are serial, being based on the Von Neumann's design. In 1959 Feynman proposed that computations could be performed at the molecular (or supramolecular) level and in a highly parallel way \cite{Feynman59}, what has led to a number of alternative proposals \cite{Conrad85}. Worth mentioning is DNA computing \cite{Adleman94}, or bacterial \cite{BactComput} and biomolecular \cite{BactComput,Benenson04} computers.
Another approach exploits the computational properties of waves in chemical excitable media (e.g. the Belousov-Zhabotinsky reaction) to solve mazes \cite{Showalter95}, perform image computation \cite{Agladze}, or design logic gates \cite{Toth}. In optics, logic gates have also been implemented using propagating solitons \cite{McLeod,Wu} and other nonlinear processes \cite{andreoni,Ceipidor}.

Localized structures in dissipative media (LS), also known as dissipative solitons, are commonplace in many spatially extended systems, such as chemical reactions, gas discharges or fluids \cite{general}. They also form in optical cavities due to the interplay between diffraction, nonlinearity, driving and dissipation. These structures have to be distinguished from conservative solitons found, for example, in propagation in fibers, for which there is a continuous family of solutions depending on their energy. Instead, dissipative LS are unique once the parameters of the system have been fixed. LS have been suggested as a potentially useful strategy for information storage \cite{FirthOPN02}. This is specially attractive in nonlinear photonics after LS have been observed in semiconductor lasers \cite{barland, Pedaci}, but the general concept of using LS to carry information is not restricted to optics.

Within this approach a LS describes one bit of information. This idea can be taken a step further and discuss the potential of LS for carrying out computations beyond mere information storage. For instance, all-optical XOR logic between the incoming bits and those stored in a cavity as LS has been proposed in \cite{Leo}. Here we propose to use excitability mediated by LS to implement three basic logic gates, namely the AND, OR, and NOT gates, providing complete logic functionality, as by combination of them one can realize any other logical operation, including the NOR and NAND universal gates. In our scheme bits are represented by a dynamical state (an excitable excursion) rather than by a stationary solution. This provides a natural reset mechanism for the gates. The way computations are performed relays on the emergent properties of the LS, independently of the microscopic details of the underlying physical system.

Excitability is a concept arising originally from biology (for example in neuroscience), and found in a large variety of nonlinear systems \cite{excitability}. A system is said to be excitable if perturbations below a certain threshold decay exponentially while perturbations above induce a large response before going back to a resting state. Roughly speaking excitability needs two ingredients: a barrier in phase space that defines the excitable threshold, and a reentry mechanism that sets the system back to the original state after a refractory time. In excitability mediated by LS the excitable threshold is automatically set by the stable manifold of the unstable (middle-branch) LS, which is the barrier one has to overcome to create a LS. The reentry mechanism is, in some cases, intrinsic to the dynamics of LS \cite{GomilaPRL,Jacobo08}, which is the case considered here. Alternatively for stationary LS, a reentry mechanism leading to excitability can be implemented by adding defect and drift in a finite system. In this case, when a super-threshold perturbation creates a LS on the defect, the drift pulls it out and drives it to the limits of the system, where the LS disappears and the system goes back to the original state. A periodic state based on this mechanism was observed in \cite{Giovanna09,Giovanna09b}, and evidence of a divergence of the period  was reported. As studied in \cite{Jacobo08}, this divergence will lead to an excitable regime. The different mechanisms discussed above makes excitability  mediated by LS highly accessible to a
wide variety of systems in optics and beyond.

\section{Nonlinear optical cavity model}

\begin{figure}[h]
\begin{center}
\includegraphics[width=0.8\textwidth]{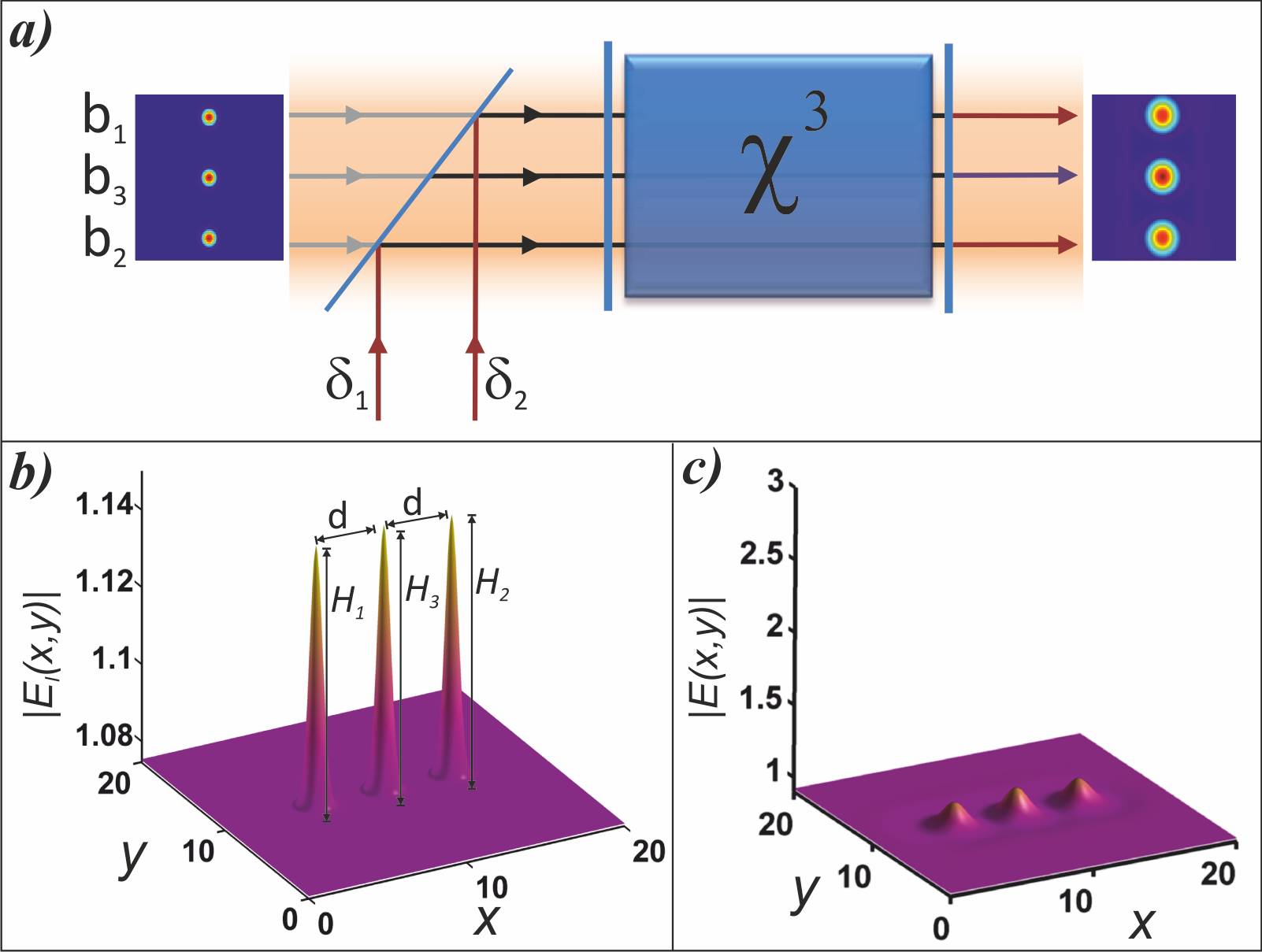}
\caption{(Color online). a) Sketch of a logic gate based on excitable LS in a nonlinear optical cavity. b) Input of the cavity $E_I(x,y)$, including the holding beam (background) and the three localized beams with intensities $H_1$, $H_2$, and $H_3$. $d$ is the distance between ports. c) Intracavity field of the resting state of the system. Adimensional units are used in all figures.}
\label{cavity}
\end{center}
\end{figure}

For illustrative purposes we consider a nonlinear optical cavity as sketched in Fig.~\ref{cavity}.
The system consist of a Kerr cavity driven by a broad holding beam. On top of this, three narrow addressing beams b$_1$, b$_2$, and b$_3$ are injected. These localized beams will facilitate excitability by allowing to tune the threshold for the size of a perturbation necessary to trigger an excitable excursion.
The positions in the transverse plane of these narrow beams define also the input and output ports of the logic gate. Alternatively the transverse plane of the device could be engineered to fix the positions of the ports. Here the bright spots created inside the cavity by b$_1$ and b$_2$ will act as input ports, while the spot at the position of b$_3$ will be the output port. The incoming bits ($ \delta_1,\delta_2$) will be superimposed to b$_1$, b$_2$.

In the mean field approximation, the slowly varying amplitude of the electric field $E(x,y,t)$ can be described by the Lugiato-Lefever model \cite{lugiato87,Jacobo08}:
\begin{equation}
\frac{\partial E}{\partial t}=-(1+i \theta)E+i
\nabla^2E+E_I+i |E^2|E,
\label{kerreq}
\end{equation}
where $\nabla^2=\partial^2/\partial x^2+\partial^2/\partial y^2$. Space and time have been rescaled with the diffraction length and cavity decay respectively, such that the coefficient in front of the Laplacian and the losses are 1. $\theta$ is the detuning between the cavity frequency and the frequency of the input field $E_I$. The sign of the cubic term corresponds the so called self-focusing case. For the AND and OR gates the input field $E_I(x,y,t)$ consists of a background $E_0$ and three Gaussian addressing beams of width $r_0$ at positions ${\bf r}_i$ $(i=1,2,3)$: $E_I(x,y,t)=E_0+\sum_{i=1}^3 [H_i+\delta_i(t)]e^{-|{\bf r}-{\bf r}_i|^2/r_0^2}$, where ${\bf r}=(x,y)$. $H_i$ is the amplitude of the addressing beams, while $\delta_i(t) (i=1,2)$ accounts for the input perturbations or bits ($\delta_3$ is always zero as the output port receives no input). $E_0$ and $H_i$ are taken real for simplicity. For the NOT gate only one addressing beam is used. Throughout this work we fix $E_0=1.0752$, $r_0=1$ and $\theta=1.45$, while $H_i$ and ${\bf r}_i$ are chosen differently for each gate. Values are given in the corresponding figure captions. Possibly, the overall input field $E$ can be implemented using an optical mask on top of a single broad beam.

Simulations are performed considering periodic boundary condition on a $512 \times 512$ mesh with $\Delta x=\Delta y=0.1875$ using a pseudospectral method as described in \cite{gomila07}. The system size $L=96$ is large enough so that boundary conditions do not affect the dynamics of the LS. Figures show only the central part of the whole system. The time step used is $\Delta t=10^{-3}$, much smaller that any relevant time scale of the system.

\begin{figure}
\begin{center}
\includegraphics[width=0.8\textwidth]{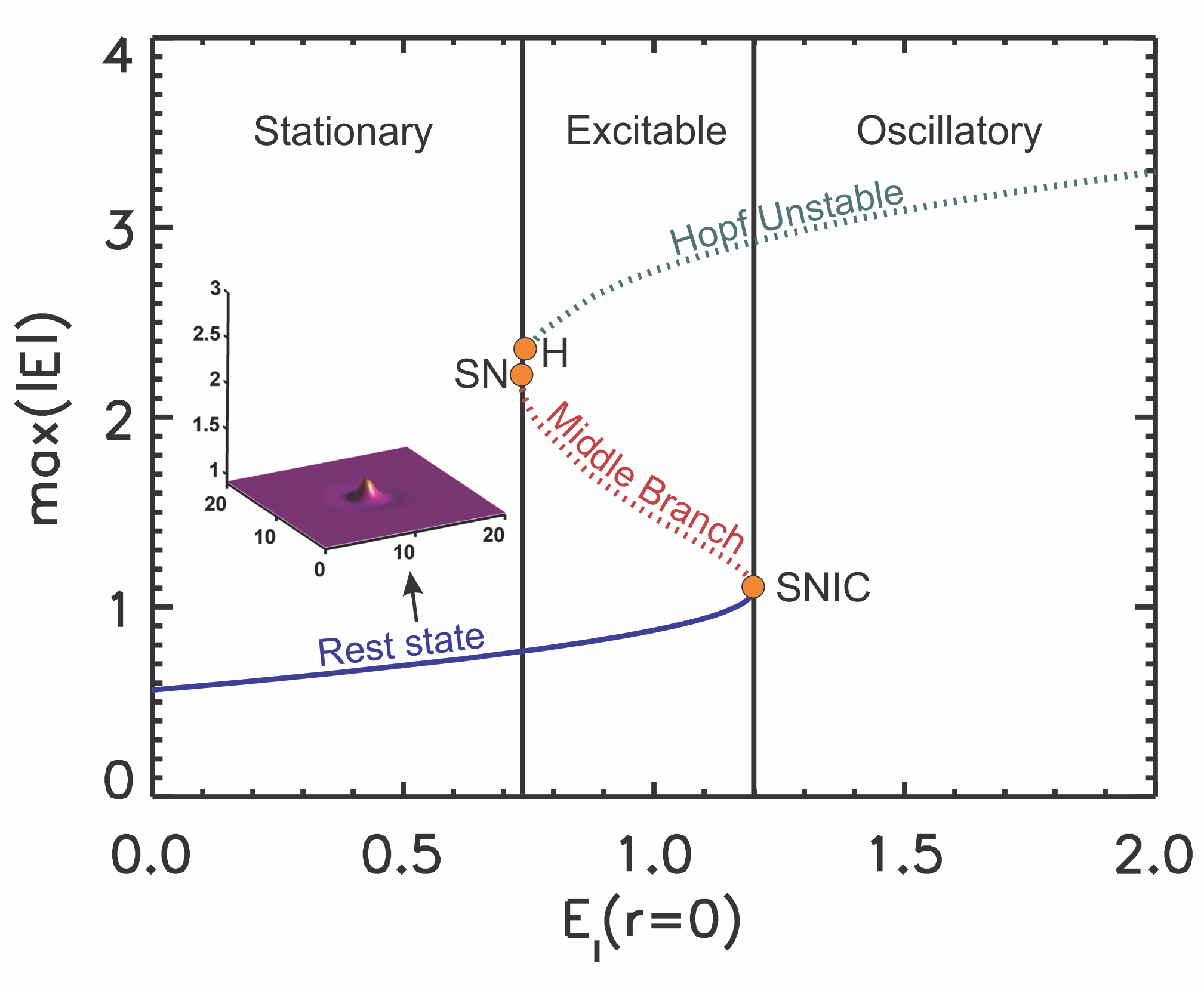}
\caption{(Color online). Bifurcation diagram of a single stationary localized solution of Eq.~(\ref{kerreq}). Solid (dotted) lines indicate stable (unstable) solutions. The SNIC bifurcation signals the frontier between excitable behavior, used for implementing AND and OR logic gates, and oscillatory behavior, used in the NOT gate.}
\label{bifdiag}
\end{center}
\end{figure}

Before analyzing how a logic gate works, lets recall the behavior of the system under a single localized addressing beam. Fig.~\ref{bifdiag} shows the bifurcation diagram of Eq.~(\ref{kerreq}) as a function of the maximum of $E_I$. The stable resting state, used as input and output ports, is represented with a solid line. This solution collides in a saddle-node on the invariant circle bifurcation (SNIC) with the unstable middle-branch LS. The middle-branch LS becomes a stable LS at the saddle-node bifurcation point (SN). For these parameter values, however, its region of stability is very narrow (barely visible in the plot), and the stationary (upper branch) becomes unstable at a very close Hopf bifurcation (H). The limit cycle arising from this bifurcation (not shown) is also almost immediately destroyed in a saddle-loop bifurcation leading to the excitable regime \cite{GomilaPRL,Jacobo08}. In this regime if the port is subject to a small perturbation the system relaxes exponentially to the resting state, while if the perturbation is larger that the excitability threshold, defined by the LS unstable middle-branch, it triggers an excitable.
The excitable excursion consists of a peak that grows to a large value until the losses stop it, and it decays back to the initial state. A remnant wave is emitted out of the center dissipating the remaining energy. For practical purposes all the region between the SN and SNIC bifurcations corresponds to a region of excitability. A full description of the parameter space where excitability appears is provided in Ref. \cite{Jacobo08}.

On the right side, excitability ends at the SNIC bifurcation, leading to an oscillatory regime. For the implementation of the AND and OR gates, relying on the excitable behavior, the system is set close but below the SNIC bifurcation. For the NOT gate, that requires an oscillatory LS, parameters are set just above the SNIC bifurcation.

\section{Logic gates}

Let us now focus on the logic operations. In our proposal a bit '1' corresponds, internally, to the presence of an excitable excursion. Then, a superthreshold perturbation at an input port (i.e. causing an excitable excursion) will be considered as a bit '1', while subthreshold (or absence of) perturbations will be considered as '0'. At the output port, the occurrence of an excitable excursion should be taken as a '1', and '0' otherwise.

\begin{figure}
\begin{center}
\includegraphics[width=0.8\textwidth]{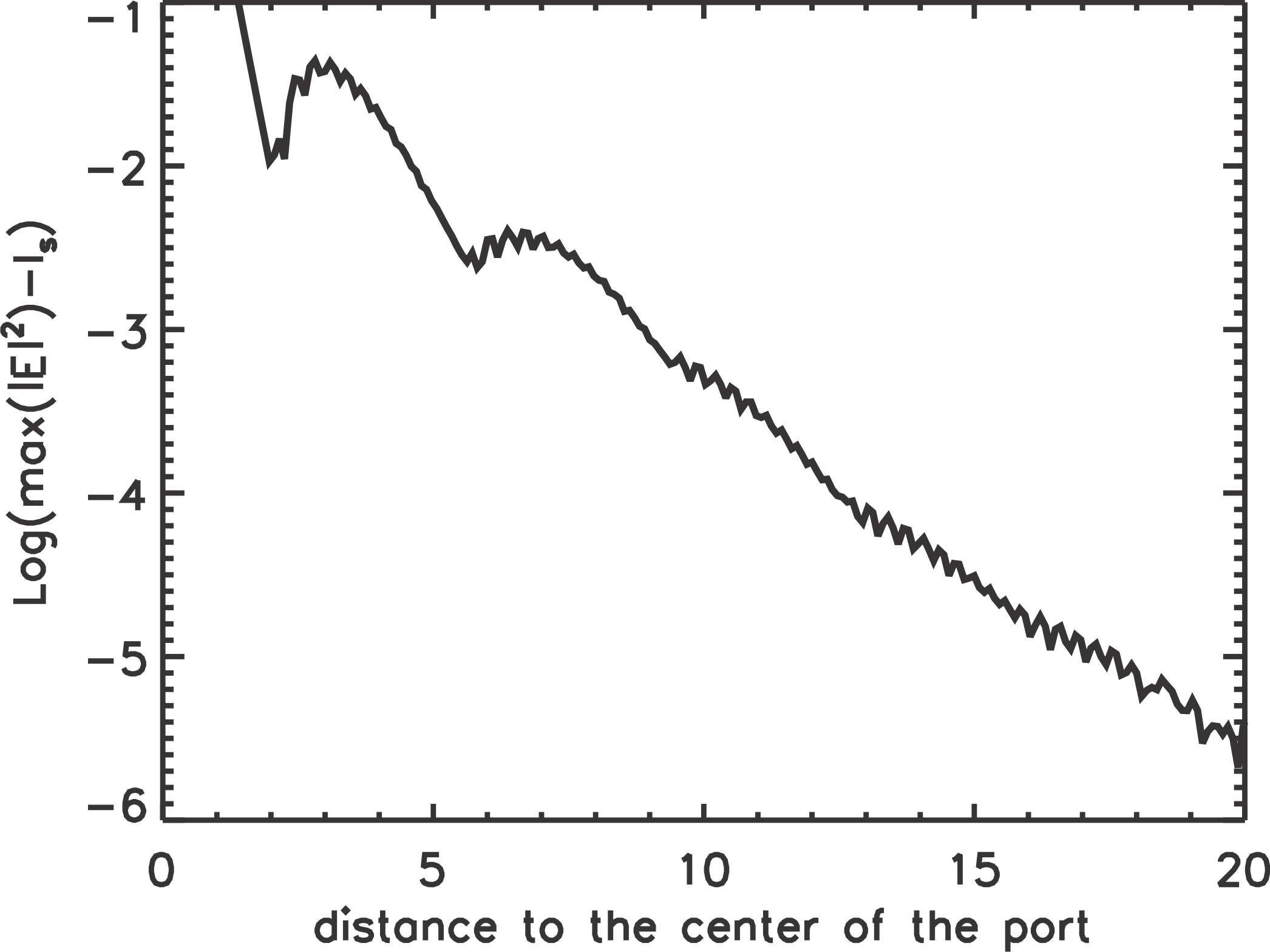}
\caption{Maximum intensity of the remnant wave as a function of the distance to the center of the port. The maximum value of the intensity at each spatial point minus the intensity of the background during a whole excitable excursion is plotted. The effect of the remnant wave decays exponentially with the distance to the port.}
\label{remwave}
\end{center}
\end{figure}

\begin{figure}
\begin{center}
\includegraphics[width=0.8\textwidth]{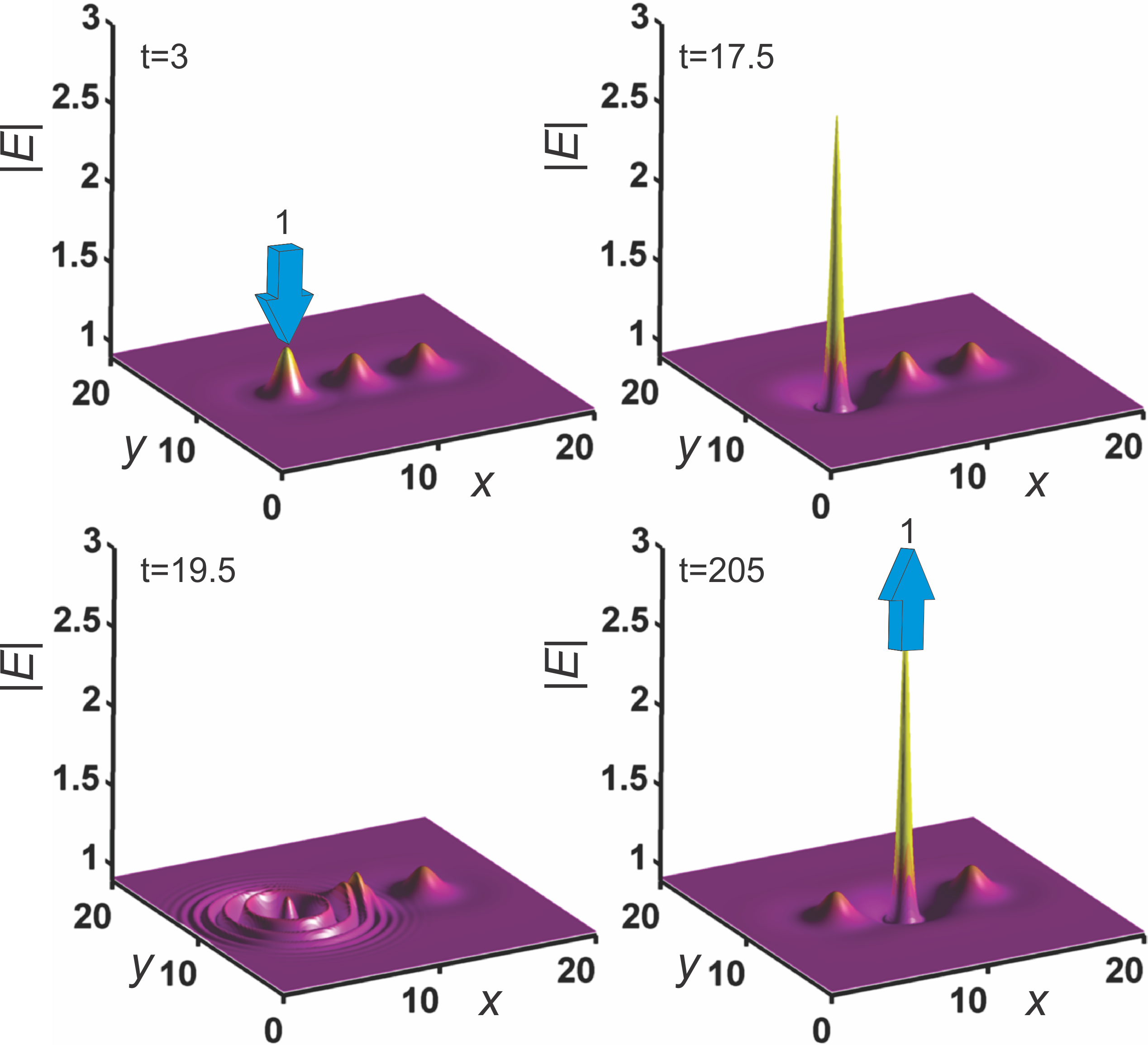}
\caption{(Color online). Time evolution of an OR logic gate with a (1,0) input.  For this case $d=8.2$,  $H_{1,2}=0.067$ for the two input (outer) ports and $H_{3}=0.0688$ for the output (central) port. The downward arrow indicates the arrival of the bit 1 at the left input port. The excitable excursion of the output port, shown by the upward arrow, gives the result (a bit 1) of the OR logic operation.}
\label{OR10}
\end{center}
\end{figure}

\begin{figure}
\begin{center}
\includegraphics[width=0.8\textwidth]{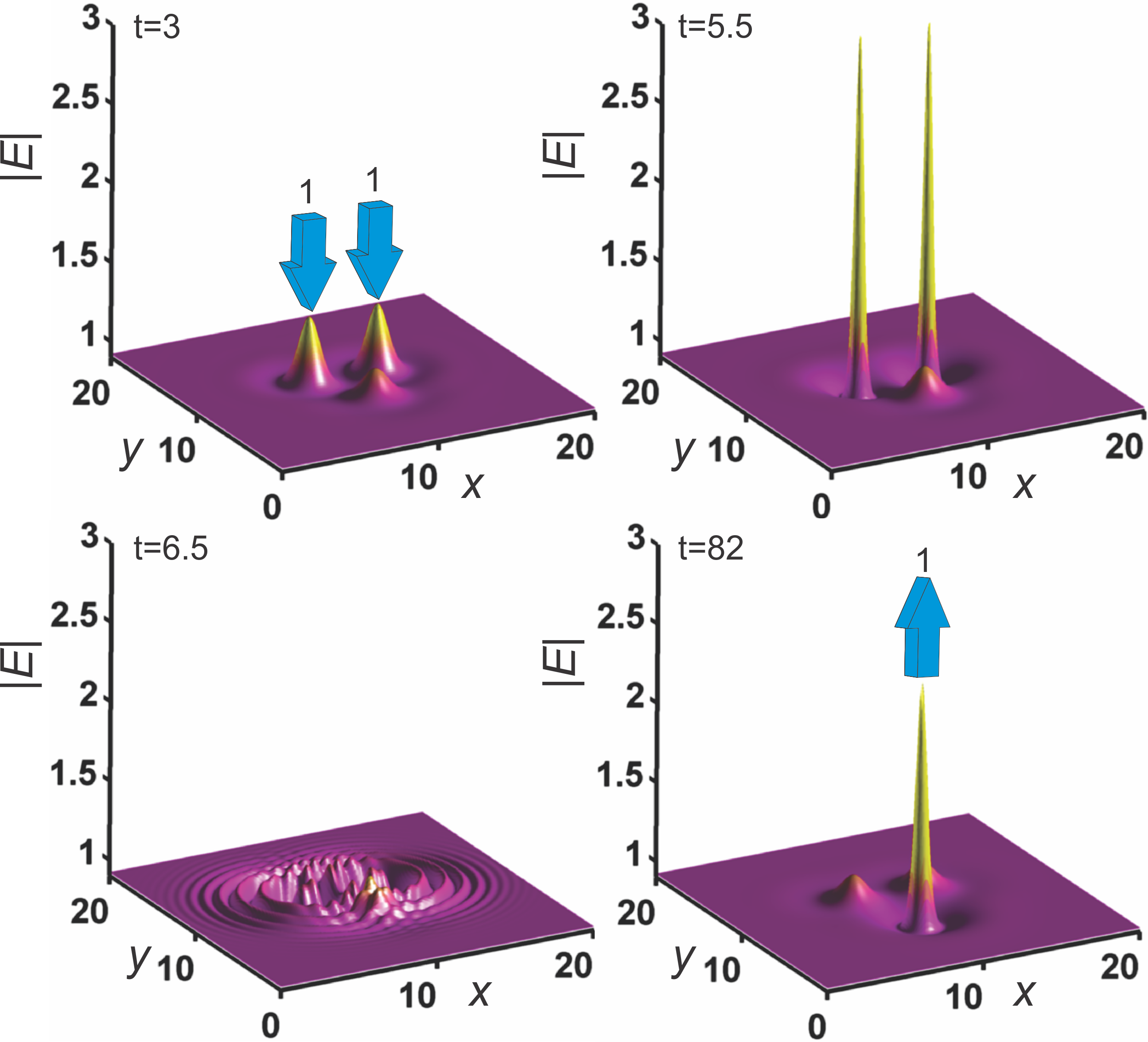}
\caption{(Color online). Time evolution of an AND logic gate with a (1,1) input. In this case  $d=8.2$,  $H_{1,2}=0.067$, and $H_{3}=0.0686$. In this case a triangular geometry is used.}
\label{AND11}
\end{center}
\end{figure}

The physical mechanism behind the computation is the interaction between ports. In particular, in our case, remnant waves that propagate energy are emitted towards the end of the excitable excursions (see Figs. \ref{OR10} and \ref{AND11}). Depending on the distance $d$ between ports, which determines the strength of the interaction, and $H_3$, that sets the sensitivity of the output port, an excitable excursion of a single input port may elicit or not an excitable excursion of the output.
As a result the truth table of the AND and OR basic gates can be reproduced by suitably tuning these parameters. Different arrangements are possible since the only relevant parameter of the geometry is the distance between ports. The interaction mediated by the remnant wave has a short range, since the system is dissipative and any perturbation decays at least exponentially with the distance and time (see Fig. \ref{remwave}).
This avoids the interference between nearby gates provided their distance is larger than the distance between the input and the output ports within a gate. To avoid the backward excitation of a input port by the output of the same gate, the excitable threshold of the input ports is set slightly higher than for the output ports. Also, the distance between the input ports is large enough to prevent their mutual excitation.

For an OR gate, $d$ and $H_3$  are such that an excitable excursion of a single input is already enough to excite the output, as shown in Fig. \ref{OR10}. The bit '1' is introduced by setting $\delta_1=0.03$ during $1$ time unit. After receiving the bit, the left input port exhibits an excitable excursion which, by the above mechanism, triggers an excitable excursion of the output port. Because of symmetry, the output would look exactly the same for the case (0,1). A double activated (1,1) input would give a very similar response, completing the truth table of the OR gate.

Similarly, AND logic can be implemented by increasing $d$ or decreasing $H_3$, such that the pulse of a single input is not enough to elicit a response of the output, but the combined action of two simultaneous excitations is (see Fig. \ref{AND11}).
To illustrate the flexibility in the geometry here we have used a triangular arrangement. This configuration has two advantages. First it is possible to do several computations sequentially in the transverse plane of the same device, instead of using a sequence of cavities, by using the output of a logic gate as one of the inputs of a contiguous one. Second, another output port can be placed symmetrically to the first one, forming a rhombus, allowing, for example, the simultaneous implementation of an AND and an OR gate. Using more complex motifs in the transverse plane more complex logical operations can be implemented in a single device. Similarly a transmission (delay) line \cite{Pedaci} with arbitrary geometry can be built connecting several excitable units.

\begin{figure}
\begin{center}
\includegraphics[width=0.8\textwidth]{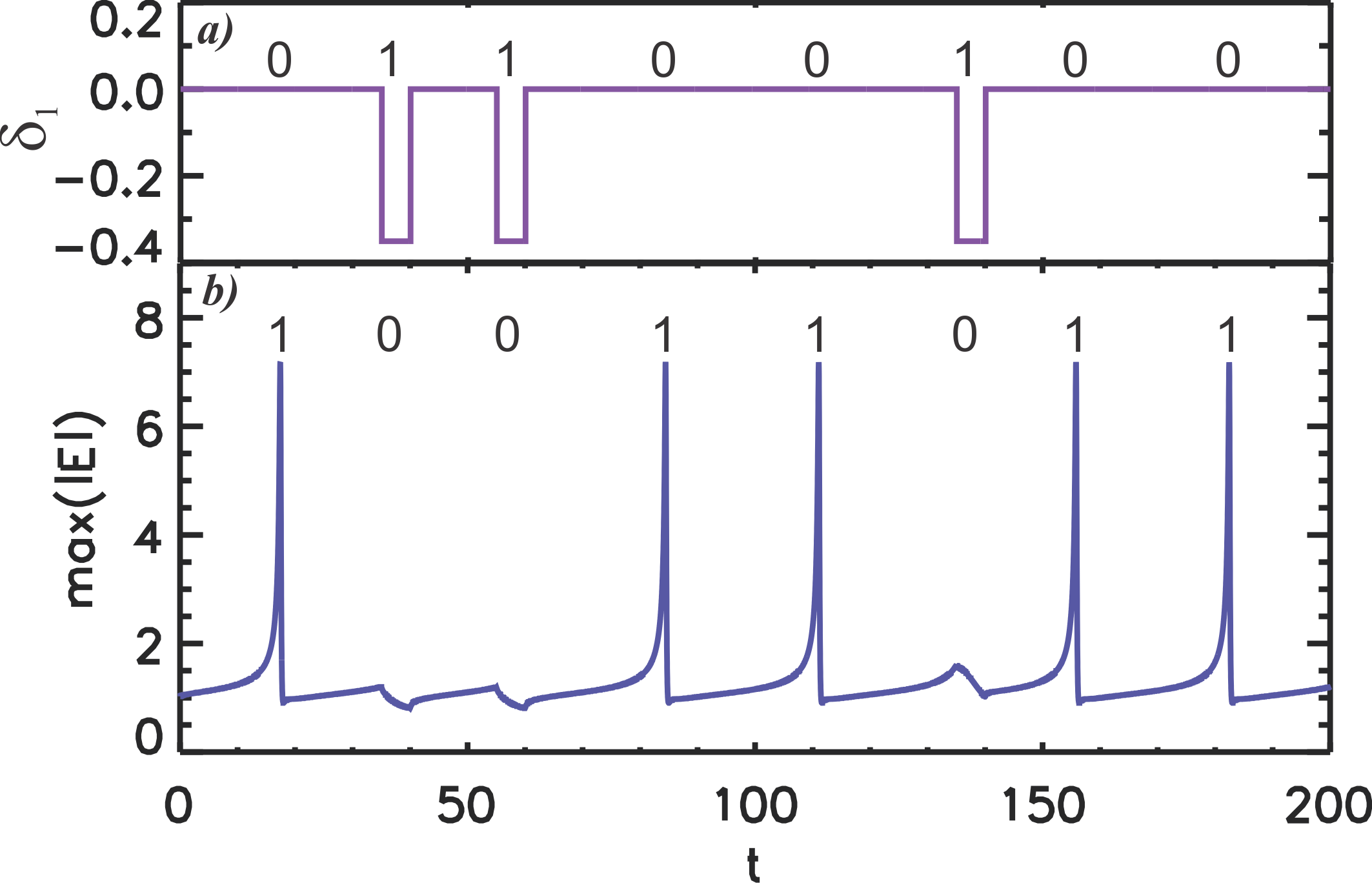}
\caption{(Color online). Time evolution of a NOT gate with an arbitrary string of input bits. a) Temporal sequence of input bits. b) Maximum of the LS showing the output of the NOT logical operation. Here $H_1=0.1233$.}
\label{NOT}
\end{center}
\end{figure}

To implement a NOT gate a different approach is needed. The NOT operation has a single input and should produce an output when no input is received. For this task we propose using a single LS in an oscillatory regime (see Fig. \ref{NOT}). The oscillations would be, on one hand, a natural clock to set the frequency of the processor, and on the other, generate a pulse at every clock step corresponding to a bit '1' if no input is received.
Then, a bit '1' at the input (with a reversed phase), introduced by setting $\delta_1=-0.03$ during $5$ time units, can set the system temporally below the oscillatory threshold, skipping one oscillation. This effectively produces a '0' at the output when a '1' is received at the input, implementing a NOT gate (see Fig. \ref{NOT}).

A key question is whether this proposal fulfills essential requirements for a logic device of any complexity to work, such as cascadability, fan-out, or logic-level restoration \cite{Miller}. On one hand, while we have chosen pulses with a square-wave time profile as input bits, these gates can work with inputs of arbitrary shape, provided its integrated energy is high enough. In particular, the energy of an excitable excursion is enough to excite several subsequent inputs, ensuring both cascadability (the output of one stage can drive the input of the next stage in a series of operations) and fan-out (the output of one stage is sufficient to drive the inputs of at least two subsequent stages). Our simulations show that an excitable pulse attenuated by a factor up to $10^{-7}$ is enough to excite the input port of a logic gate in another cavity. On the other hand, logic-level restoration is provided by the intrinsic nonlinear dynamics of the system, since, provided the input perturbation is above the threshold, the excitable excursion takes place independently of its details. Thus, any signal is automatically restored independently of its level of degradation at the input. Besides, this scheme is quite robust to spontaneous emission, since spatially uncorrelated fluctuations have a very small effect on LS which are relatively broad objects \cite{noise}.

Considering the low energy requirements for optical transmission, an optical transistor would already be useful even if its performance is only similar to that one of the electronic counterparts, due to the advantage of avoiding the conversion of information from the optical to the electronic domain.

A possible issue of our proposal is isochronicity, as different logical operations may have different response times. These differences however, may not necessarily prevent practical operation since they might fall within acceptable tolerances. Otherwise a buffer memory could be implemented to synchronize different outputs. This could be done using the same LS in the bistable regime. Finally, we should note that although we have used coherent localized beams, it has been shown that LS can also be switched on and off incoherently \cite{Giovanna09,Lugiato,Leo}.

\section{Conclusions}

Summarizing, we have analyzed the possibility of creating logic gates using the dynamics of LS. In particular, we have shown how to construct both an AND and an OR gate by appropriately coupling three excitable ports, and a NOT gate using a single LS in an oscillatory regime. It is important to stress that this proof of concept is independent of the specific model considered here, and that logic operations could be realized in any system displaying LS, even beyond optics. More generally, this full logical functionality is a way to show that a very large class of pattern forming systems are Turing universal \cite{winningways}. From a practical point of view vertical cavity surface emitting lasers (VCSELs) are particularly interesting, as in addition to its integrability with semiconductor technology, oscillatory LS have already been experimentally observed \cite{Thorsten,Giovanna09b}, and evidence of excitable behavior has been theoretically observed in lasers with frequency selective feedback \cite{Pavel}. Evidence of a SNIC bifurcation for stationary LS leading to excitability in the presence of a defect and drift has also been reported in \cite{Giovanna09,Giovanna09b}. Moreover, the temporal response in VCSELs can be as high as 10 $GHz$, possibly allowing computations in the $GHz$ range. Oscillating, and possibly excitable, polariton LS have also been recently reported in semiconductor microcavities \cite{Egorov}.

This computational method has also potential advantages with respect to conventional electronics. First, it is intrinsically parallel (several logic gates can be simultaneously operated in a single device), allowing for instance dense optical interconnects to shine bits directly on a battery of input ports on the same broad area device. Second, it is reconfigurable by changing only the ''mask`` defining the position and intensity of the localized beams. Since this can be done all-optically, logic circuits can be reprogrammed in real time, at rates of the order of the clock frequency. Otherwise, the same role of the localized beams can be played by defects of the material, that can be engineered to created hard-wired gates.

We acknowledge financial support from MICINN (Spain) and FEDER (EU) through Grants FIS2007-60327(FISICOS) and TEC2009-14101 (DeCoDicA). DG acknowledge financial support from CSIC (Spain) through grant number 201050I016.


\begin{thebibliography}{29}
\expandafter\ifx\csname natexlab\endcsname\relax\def\natexlab#1{#1}\fi
\expandafter\ifx\csname bibnamefont\endcsname\relax
  \def\bibnamefont#1{#1}\fi
\expandafter\ifx\csname bibfnamefont\endcsname\relax
  \def\bibfnamefont#1{#1}\fi
\expandafter\ifx\csname citenamefont\endcsname\relax
  \def\citenamefont#1{#1}\fi
\expandafter\ifx\csname url\endcsname\relax
  \def\url#1{\texttt{#1}}\fi
\expandafter\ifx\csname urlprefix\endcsname\relax\def\urlprefix{URL }\fi
\providecommand{\bibinfo}[2]{#2}
\providecommand{\eprint}[2][]{\url{#2}}


\bibitem{Feynman59}
\bibinfo{author}{\bibfnamefont{R.~P.} \bibnamefont{Feynman}}, in
  \emph{\bibinfo{booktitle}{Miniaturization}}, edited by
  \bibinfo{editor}{\bibfnamefont{D.}~\bibnamefont{Gilbert}}
  (\bibinfo{publisher}{Reinhold}, \bibinfo{year}{1961}), pp.
  \bibinfo{pages}{282--296}.

\bibitem{Conrad85}
\bibinfo{author}{\bibfnamefont{M.}~\bibnamefont{Conrad}},
  \bibinfo{journal}{Commun. ACM} \textbf{\bibinfo{volume}{28}},
  \bibinfo{pages}{464} (\bibinfo{year}{1985});
  \bibinfo{author}{\bibnamefont{A.~Hjelmfelt}, \bibfnamefont{E.~D. Weinberger}}
  \bibnamefont{and} \bibinfo{author}{\bibfnamefont{J.}~\bibnamefont{Ross}},
  \bibinfo{journal}{Proc. Natl. Acad. Sci. USA} \textbf{\bibinfo{volume}{88}},
  \bibinfo{pages}{10983} (\bibinfo{year}{1991}).

\bibitem{Adleman94}
\bibinfo{author}{\bibfnamefont{L.~M.} \bibnamefont{Adleman}},
  \bibinfo{journal}{Science} \textbf{\bibinfo{volume}{266}},
  \bibinfo{pages}{1021} (\bibinfo{year}{1994});

\bibinfo{author}{\bibfnamefont{R.~J.} \bibnamefont{Lipton}},
  \bibinfo{journal}{Science} \textbf{\bibinfo{volume}{268}},
  \bibinfo{pages}{542} (\bibinfo{year}{1995});
\bibinfo{author}{\bibfnamefont{P.~W.~K.} \bibnamefont{Rothemund}},
  \bibinfo{journal}{Nature} \textbf{\bibinfo{volume}{440}},
  \bibinfo{pages}{297} (\bibinfo{year}{2006});
\bibinfo{author}{\bibfnamefont{E.}~\bibnamefont{Shapiro}} \bibnamefont{and}
  \bibinfo{author}{\bibfnamefont{Y.}~\bibnamefont{Benenson}},
  \bibinfo{journal}{Sci. Am.} \textbf{\bibinfo{volume}{294}},
  \bibinfo{pages}{44} (\bibinfo{year}{2006}).

\bibitem{BactComput}
\bibinfo{author}{\bibfnamefont{J.}~\bibnamefont{Baumgardner }}
  {\it et al.}
  \bibinfo{journal}{J. Biol. Eng.} \textbf{\bibinfo{volume}{3}},
  \bibinfo{pages}{11} (\bibinfo{year}{2009}).

\bibitem{Benenson04}
\bibinfo{author}{\bibfnamefont{Y.}~\bibnamefont{Benenson}}
  {\it et al.},
  \bibinfo{journal}{Nature} \textbf{\bibinfo{volume}{429}},
  \bibinfo{pages}{423} (\bibinfo{year}{2004}).

\bibitem{Showalter95}
\bibinfo{author}{\bibfnamefont{O.}~\bibnamefont{Steinbock}},
  \bibinfo{author}{\bibfnamefont{A.}~\bibnamefont{T\'oth}}, \bibnamefont{and}
  \bibinfo{author}{\bibfnamefont{K.}~\bibnamefont{Showalter}},
  \bibinfo{journal}{Science} \textbf{\bibinfo{volume}{267}},
  \bibinfo{pages}{868} (\bibinfo{year}{1995}).

\bibitem{Agladze}
\bibinfo{author}{\bibfnamefont{L.}~\bibnamefont{Kuhnert}},
  \bibinfo{author}{\bibfnamefont{K.~I.} \bibnamefont{Agladze}},
  \bibnamefont{and} \bibinfo{author}{\bibfnamefont{V.~I.}
  \bibnamefont{Krinsky}}, \bibinfo{journal}{Nature}
  \textbf{\bibinfo{volume}{337}}, \bibinfo{pages}{244} (\bibinfo{year}{1989}).

\bibitem{Toth}
\bibinfo{author}{\bibfnamefont{A.}~\bibnamefont{Toth}} \bibnamefont{and}
  \bibinfo{author}{\bibfnamefont{K.}~\bibnamefont{Showalter}},
  \bibinfo{journal}{J. Chem. Phys.} \textbf{\bibinfo{volume}{103}},
  \bibinfo{pages}{2058} (\bibinfo{year}{1995});
\bibinfo{author}{\bibfnamefont{O.}~\bibnamefont{Steinbock}},
  \bibinfo{author}{\bibfnamefont{P.}~\bibnamefont{Kettunen}}, \bibnamefont{and}
  \bibinfo{author}{\bibfnamefont{K.}~\bibnamefont{Showalter}},
  \bibinfo{journal}{J. Phys. Chem.} \textbf{\bibinfo{volume}{100}},
  \bibinfo{pages}{18970} (\bibinfo{year}{1996});
\bibinfo{author}{\bibfnamefont{J.}~\bibnamefont{Gorecka}} \bibnamefont{and}
  \bibinfo{author}{\bibfnamefont{J.}~\bibnamefont{Gorecki}},
  \bibinfo{journal}{J. Chem. Phys.} \textbf{\bibinfo{volume}{124}},
  \bibinfo{pages}{084101} (\bibinfo{year}{2006}).


 \bibitem{McLeod}
\bibinfo{author}{\bibfnamefont{R.}~\bibnamefont{McLeod}},
\bibinfo{author}{\bibfnamefont{K.}~\bibnamefont{Wagner}}, and
\bibinfo{author}{\bibfnamefont{S.}~\bibnamefont{Blair}},
\bibinfo{journal}{Phys. Rev. A}
  \textbf{\bibinfo{volume}{52}}, \bibinfo{pages}{3254} (\bibinfo{year}{1995}).

 \bibitem{Wu}
\bibinfo{author}{\bibfnamefont{Y.-D.}~\bibnamefont{Wu}},
\bibinfo{journal}{Opt. Express}
  \textbf{\bibinfo{volume}{14}}, \bibinfo{pages}{4005} (\bibinfo{year}{2006}).

\bibitem{andreoni}
\bibinfo{author}{\bibfnamefont{A.}~\bibnamefont{Andreoni}},
\bibinfo{author}{\bibfnamefont{M.}~\bibnamefont{Bondani}}, and
\bibinfo{author}{\bibfnamefont{A.C.}~\bibnamefont{Potenza}},
\bibinfo{journal}{Rev. Sci. Instrum.}
  \textbf{\bibinfo{volume}{72}}, \bibinfo{pages}{2525} (\bibinfo{year}{2001}).

  \bibitem{Ceipidor}
\bibinfo{author}{\bibfnamefont{L.}~\bibnamefont{Biader Ceipidor}},
\bibinfo{author}{\bibfnamefont{A.}~\bibnamefont{Bosco}}, and
\bibinfo{author}{\bibfnamefont{E.}~\bibnamefont{Fazio}},
\bibinfo{journal}{IEEE - J. Lightwave Technol.}
  \textbf{\bibinfo{volume}{26}}, \bibinfo{pages}{373} (\bibinfo{year}{2008}).


\bibitem{general}
\bibinfo{author}{\bibfnamefont{N.}~\bibnamefont{Akhmediev}} and
\bibinfo{author}{\bibfnamefont{A.}~\bibnamefont{Ankiewicz}} (eds.)
  \emph{\bibinfo{title}{{Dissipative Solitons}}},
  (\bibinfo{publisher}{Springer}, \bibinfo{address}{New
  York}, \bibinfo{year}{2005});
\bibinfo{author}{\bibfnamefont{H.-G.} \bibnamefont{Purwins}},
\bibinfo{author}{\bibfnamefont{H.U.} \bibnamefont{B\"odeker}},
\bibnamefont{and}
  \bibinfo{author}{\bibfnamefont{Sh.} \bibnamefont{Amiranashvili}},
  \bibinfo{journal}{Adv. Phys.} \textbf{\bibinfo{volume}{59}},
  \bibinfo{pages}{485} (\bibinfo{year}{2010}).


\bibitem{FirthOPN02}
\bibinfo{author}{\bibfnamefont{W.~J.} \bibnamefont{Firth}} \bibnamefont{and}
  \bibinfo{author}{\bibfnamefont{C.~O.} \bibnamefont{Weiss}},
  \bibinfo{journal}{Opt. Photonic News} \textbf{\bibinfo{volume}{13}},
  \bibinfo{pages}{54} (\bibinfo{year}{2002});
\bibinfo{author}{\bibfnamefont{P.}~\bibnamefont{Coullet}},
  \bibinfo{author}{\bibfnamefont{C.}~\bibnamefont{Riera}}, \bibnamefont{and}
  \bibinfo{author}{\bibfnamefont{C.}~\bibnamefont{Tresser}},
  \bibinfo{journal}{Chaos} \textbf{\bibinfo{volume}{14}}, \bibinfo{pages}{193}
  (\bibinfo{year}{2004}).

\bibitem{barland}
\bibinfo{author}{\bibfnamefont{S.}~\bibnamefont{Barland}} {\it et al.},
\bibinfo{journal}{Nature (London)}
  \textbf{\bibinfo{volume}{419}}, \bibinfo{pages}{699} (\bibinfo{year}{2002}).

\bibitem{Pedaci}
\bibinfo{author}{\bibfnamefont{F.}~\bibnamefont{Pedaci}}
{\it et al.},
 \bibinfo{journal}{Appl. Phys. Lett.}
  \textbf{\bibinfo{volume}{92}}, \bibinfo{pages}{011101}
  (\bibinfo{year}{2008}).

\bibitem{Leo}
\bibinfo{author}{\bibfnamefont{F.}~\bibnamefont{Leo}}
  {\it et al.},
  \bibinfo{journal}{Nature Photon.} \textbf{\bibinfo{volume}{4}},
  \bibinfo{pages}{471} (\bibinfo{year}{2010}).

\bibitem{excitability}
\bibinfo{author}{\bibfnamefont{A.}~\bibnamefont{Winfree}},
  \emph{\bibinfo{title}{{The Geometry of Biological Time}}},
  (\bibinfo{publisher}{Springer-Verlag}, \bibinfo{address}{New
  York}, \bibinfo{year}{2001}), \bibinfo{edition}{2nd} ed.

\bibitem{GomilaPRL}
\bibinfo{author}{\bibfnamefont{D.}~\bibnamefont{Gomila}},
  \bibinfo{author}{\bibfnamefont{M.~A.} \bibnamefont{Mat{\'{\i}}as}},
  \bibnamefont{and} \bibinfo{author}{\bibfnamefont{P.}~\bibnamefont{Colet}},
  \bibinfo{journal}{Phys. Rev. Lett.} \textbf{\bibinfo{volume}{94}},
  \bibinfo{pages}{063905} (\bibinfo{year}{2005}).

\bibitem{Jacobo08}
\bibinfo{author}{\bibfnamefont{A.}~\bibnamefont{Jacobo}} {\it et al.},
  \bibinfo{journal}{Phys. Rev. A} \textbf{\bibinfo{volume}{78}},
  \bibinfo{pages}{053821} (\bibinfo{year}{2008}).

\bibitem{Giovanna09}
\bibinfo{author}{\bibfnamefont{E.}~\bibnamefont{Caboche}}
  {\it et al.},
  \bibinfo{journal}{Phys. Rev. A} \textbf{\bibinfo{volume}{80}},
  \bibinfo{pages}{053814} (\bibinfo{year}{2009}{\natexlab{a}}).

\bibitem{Giovanna09b}
\bibinfo{author}{\bibfnamefont{E.}~\bibnamefont{Caboche}}
  {\it et al.},
  \bibinfo{journal}{Phys. Rev. Lett.} \textbf{\bibinfo{volume}{102}},
  \bibinfo{pages}{163901} (\bibinfo{year}{2009}{\natexlab{b}}).

\bibitem{lugiato87}
\bibinfo{author}{\bibfnamefont{L.~A.} \bibnamefont{Lugiato}} \bibnamefont{and}
  \bibinfo{author}{\bibfnamefont{R.}~\bibnamefont{Lefever}},
  \bibinfo{journal}{Phys. Rev. Lett.} \textbf{\bibinfo{volume}{58}},
  \bibinfo{pages}{2209} (\bibinfo{year}{1987}).


  \bibitem{gomila07}
\bibinfo{author}{\bibfnamefont{D.} \bibnamefont{Gomila}}
{\it et al.},
  \bibinfo{journal}{Phys. Rev. E} \textbf{\bibinfo{volume}{75}},
  \bibinfo{pages}{026217} (\bibinfo{year}{2007}).


\bibitem{Miller}
\bibinfo{author}{\bibfnamefont{D.}~\bibnamefont{Miller}},
  \bibinfo{journal}{Nature Photon.} \textbf{\bibinfo{volume}{4}},
  \bibinfo{pages}{3} (\bibinfo{year}{2010}).

\bibitem{noise} A. Jacobo {\it et al.},  Eur. Phys. J. D {\bf 59}, 37 (2010).


\bibitem{Lugiato}
\bibinfo{author}{\bibfnamefont{K.}~\bibnamefont{Aghdami}}
{\it et al.},
  \bibinfo{journal}{European Phys. J. D} \textbf{\bibinfo{volume}{47}},
  \bibinfo{pages}{447} (\bibinfo{year}{2008}).

\bibitem{Thorsten}
\bibinfo{author}{\bibfnamefont{T.}~\bibnamefont{Ackemann}}
  {\it et al.},
  \bibinfo{journal}{preprint}  (\bibinfo{year}{2010}).

\bibitem{Pavel}
\bibinfo{author}{\bibfnamefont{P.V.}~\bibnamefont{Paulau}}
{\it et al.},
  \bibinfo{journal}{Phys. Rev. A} \textbf{\bibinfo{volume}{80}},
  \bibinfo{pages}{023808} (\bibinfo{year}{2009}).


\bibitem{winningways}
\bibinfo{author}{\bibfnamefont{E.R.}~\bibnamefont{Berlekamp}},
  \bibinfo{author}{\bibfnamefont{J.H.}~\bibnamefont{Conway}}, \bibnamefont{and}
  \bibinfo{author}{\bibfnamefont{R.K.}~\bibnamefont{Guy}},
  \emph{\bibinfo{title}{{Winning Ways, Vol. 2}}},
  (\bibinfo{publisher}{Academic Press}, \bibinfo{address}{London}, \bibinfo{year}{1982}).

\bibitem{Egorov}
\bibinfo{author}{\bibfnamefont{O.A.}~\bibnamefont{Egorov}},
  \bibinfo{author}{\bibfnamefont{D.V.}~\bibnamefont{Skryabin}}, \bibnamefont{and}
  \bibinfo{author}{\bibfnamefont{F.}~\bibnamefont{Lederer}},
  \bibinfo{journal}{Phys. Rev. B} \textbf{\bibinfo{volume}{82}},
  \bibinfo{pages}{165326} (\bibinfo{year}{2010}).


\end{thebibliography}
\end{document}